\documentclass[sigconf, natbib=false, nonacm]{acmart}
\usepackage[datamodel=acmdatamodel, style=acmnumeric, backend=biber]{biblatex}
\usepackage{amsmath, amsfonts}

\usepackage{graphicx}
\usepackage{subcaption}

\usepackage[nameinlink]{cleveref}

\usepackage[inline]{enumitem}
\setlist*[itemize]{labelindent=10pt, itemindent=0pt, leftmargin=*}

\usepackage{booktabs}
\usepackage{multirow}
\usepackage{tabularx}

\usepackage{pgfplots}
\pgfplotsset{compat=1.18}
\usepgfplotslibrary{statistics}

\usepackage{csquotes}
\usepackage{textcomp}

\usepackage{balance}

\begin{document}

\title[Model Parallelism on Distributed Infrastructure: A Literature Review]{Model Parallelism on Distributed Infrastructure: A Literature Review from Theory to LLM Case-Studies}

\author{Felix Brakel}
\orcid{0009-0004-1730-3005}
\affiliation{%
	\institution{Informatics Institute, University of Amsterdam}
	\city{Amsterdam}
	\country{The Netherlands}}
\affiliation{%
	\institution{Vrije Universiteit Amsterdam}
	\city{Amsterdam}
	\country{The Netherlands}}
\email{felix.brakel@student.uva.nl}

\author{Uraz Odyurt}
\orcid{0000-0003-1094-0234}
\affiliation{%
	\institution{High-Energy Physics, Radboud University}
	\city{Nijmegen}
	\country{The Netherlands}}
\affiliation{%
	\institution{National Institute for Subatomic Physics (Nikhef)}
	\city{Amsterdam}
	\country{The Netherlands}}
\email{uodyurt@nikhef.nl}

\author{Ana-Lucia Varbanescu}
\orcid{0000-0002-4932-1900}
\affiliation{%
	\institution{Computer Architecture for Embedded Systems, University of Twente}
	\city{Enschede}
	\country{The Netherlands}}
\affiliation{%
	\institution{Informatics Institute, University of Amsterdam}
	\city{Amsterdam}
	\country{The Netherlands}}
\email{a.l.varbanescu@utwente.nl}

\renewcommand{\shortauthors}{F. Brakel et al.}

\begin{abstract}
Neural networks have become a cornerstone of machine learning. As the trend for these to get more and more complex continues, so does the underlying hardware and software infrastructure for training and deployment. In this survey we answer three research questions: \emph{\enquote{What types of model parallelism exist?}}, \emph{\enquote{What are the challenges of model parallelism?}}, and \emph{\enquote{What is a modern use-case of model parallelism?}} We answer the first question by looking at how neural networks can be parallelised and expressing these as operator graphs while exploring the available dimensions. The dimensions along which neural networks can be parallelised are \emph{intra-operator} and \emph{inter-operator}. We answer the second question by collecting and listing both implementation challenges for the types of parallelism, as well as the problem of optimally partitioning the operator graph. We answer the last question by collecting and listing how parallelism is applied in modern multi-billion parameter transformer networks, to the extend that this is possible with the limited information shared about these networks.
\end{abstract}

%
%


\maketitle


\section{Introduction}
\label{sec:introduction}
Neural networks have become a cornerstone in machine learning, offering solutions for complex prediction tasks. As these networks grow in complexity, both computational requirements and memory footprint for training and inference, increase proportionally.

The increase in computational requirements is due to the greater number of operations needed to perform tasks like forward and backward passes during training. More complex models often have more layers, more neurons, or more sophisticated architectures, all of which contribute to an increased number of mathematical operations. Similarly, the memory footprint increases because more complex models require more parameters, and each parameter needs to be stored in memory. Additionally, intermediate values generated during computation also consume memory, and their number grows with the complexity of the model.

One way to continue meeting these computational demands is through model parallelism: by partitioning the model the workload can be spread out over multiple devices. However, the data-intensity of neural network workloads makes this non-trivial. Both the parameters and the data flowing through the network are of considerable size and when distributing the neural network over multiple devices this data now has to be send over an interconnect such as a high-speed NVLink bridge or a regular Ethernet connection.

Compared to fetching of data from memory, these interconnects pose serious bandwidth limitations. Even when only considering a single server, where devices can send data over NVLink, the bandwidth is already a factor two below that of the A100's DRAM~\cite{Choquette:2021:NATC}. Often however, we are trying to scale even beyond this to multiple nodes, where communication between nodes passes over a comparably glacial network built on for example, Ethernet. 

Model parallelism then has the potential to meet the ever-growing demands computational demands of neural networks. In this survey we aim to provide a view on model parallelism by answering the following questions:
\begin{enumerate}
    \item \emph{What types of model parallelism exist?}
    \item \emph{What are the challenges of model parallelism?}
    \item \emph{What is a modern use-case of model parallelism?}
\end{enumerate}

\paragraph{Outline}
\Cref{sec:study_design} defines the considered constraints in our study design, followed by a detailed background on model parallelism in \Cref{sec:model_parallelism}. \Cref{sec:challenges} covers the collected challenges, while \Cref{sec:use_cases} delves into the details of collected use-cases. Following relevant discussions in \Cref{sec:discussion}, we conclude in \Cref{sec:conclusion} by revisiting our initial research questions.

\section{Study design}
\label{sec:study_design}
This study consists of two distinct phases. In order to provide a theoretical framework for neural network workloads and to tackle the first research question \emph{\enquote{What types of model parallelism exist?}}, the first phase consists of a study in Deep Neural Network (DNN) auto-parallelisation. DNN auto-parallelisation formulates model parallelism in a form suitable for search algorithms. The literature collection process for this phase was done using a snowballing approach, with the 2023 survey by~\cite{Liang:2023:SAPL} as seed. The available papers were filtered according to the following criteria:
\begin{enumerate}
    \item Code state: Available - 22 papers left after filtering
    \item Code state: Actively maintained - 8 papers
    \item Neural network training type: Fully automated - 7 papers
    \item Compatibility with existing file formats - 6 papers
\end{enumerate}

\begin{figure}[htbp]
    \centering        
    \includegraphics[width=\linewidth]{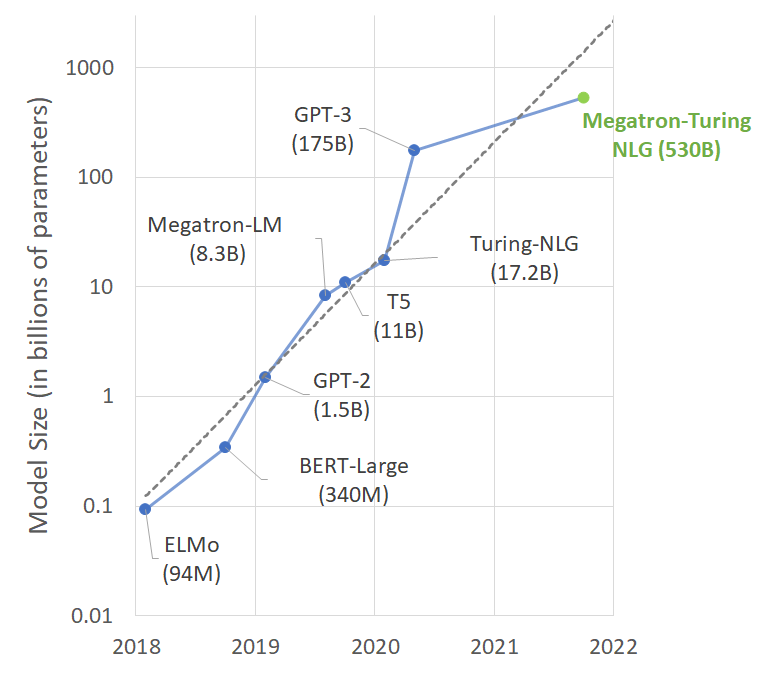}
    \caption{Megatron-NLG compared to other large language models (source:~\cite{Smith:2022:UDSM}).}
    \label{fig:transformer_size}
\end{figure}

In order to answer the second and the third research questions, the second phase consists of a study into how model parallelism is used in modern Transformer networks. This phase too was performed following a snowballing approach. The seeds for the second phase are the models from \Cref{fig:transformer_size} (taken from~\cite{Smith:2022:UDSM}) and \Cref{tab:transformer_size} (taken from~\cite{Chowdhery:2023:PALM}). 
\begin{table}[htbp]
    \centering
    \caption{Comparison of different Transformer models (source:~\cite{Chowdhery:2023:PALM}).}
    \begin{tabular}{lrlr}
    	\toprule
    	Model 				& \# Parameters 		& Hardware 			& Utilisation \\
    	\midrule
    	GPT-3 				& 175B 					& V100 				& 21.3\% \\
    	Gopher 				& 280B 					& 4096 TPU v3 		& 32.5\% \\
    	Megatron-Turing 	& 530B 					& 2240 A100 		& 30.2\% \\
    	PaLM 				& 540B 					& 6144 TPUv4 		& 46.2\% \\
    	\bottomrule
    \end{tabular}
    \label{tab:transformer_size}
\end{table}

This phase resulted in the collection of four papers on the Megatron family of models, one about Gopher, two about PaLM and two about GPT, listed further ahead in \Cref{tab:transformer_results}. Sadly, details about the implementations of these models, as it pertains to model parallelism, are scarce. Thus, we filter the papers on the availability of these details, after which, we are left with the mentioned four papers on Megatron, the paper about Gopher and one paper about PaLM.

\section{Model parallelism}
\label{sec:model_parallelism}
Model parallelism in neural networks is characterised by partitioning the model itself and distributing the partitions over multiple compute devices. This approach offers potential benefits, both in model throughput and in lowering per-device memory requirements. To further define what model parallelism is, we first offer a framework for reasoning about neural networks from a computational perspective. We provide a background on model parallelism as neural networks operating in the Single Instruction Multiple Data (SIMD) form. Following that, we answer the first research question, i.e., \emph{\enquote{What types of model parallelism exist?}}, considering theoretical and implementation-related perspectives.

\subsection{Background}
\label{subsec:background}
In machine learning we distinguish between two phases: training and inference. At training-time we train a model on a set of data called the learning set. At inference-time we task the trained model with making predictions on new, unseen, data.

One of such models is a neural network and for complex prediction tasks they dominate the state of the art. We will give a more detailed explanation of a neural network in \Cref{subsec:neural_network_workloads}  but for now a conceptual explanation will suffice.

Neural Networks (NNs) are, as the name suggests, made up of artificial neurons. The neurons are organised in layers. Neurons in a layer all perform the same operation on their input data and thus these layers are also referred to as \emph{operators}. Layers have weighted connections between and it is by adjusting these weights during training time that the NN is able to \emph{learn}. These weights are referred to as model's \emph{parameters}.

\subsubsection{Scaling up networks}
\begin{figure}[htbp]
    \centering
    \includegraphics[width=0.65\linewidth]{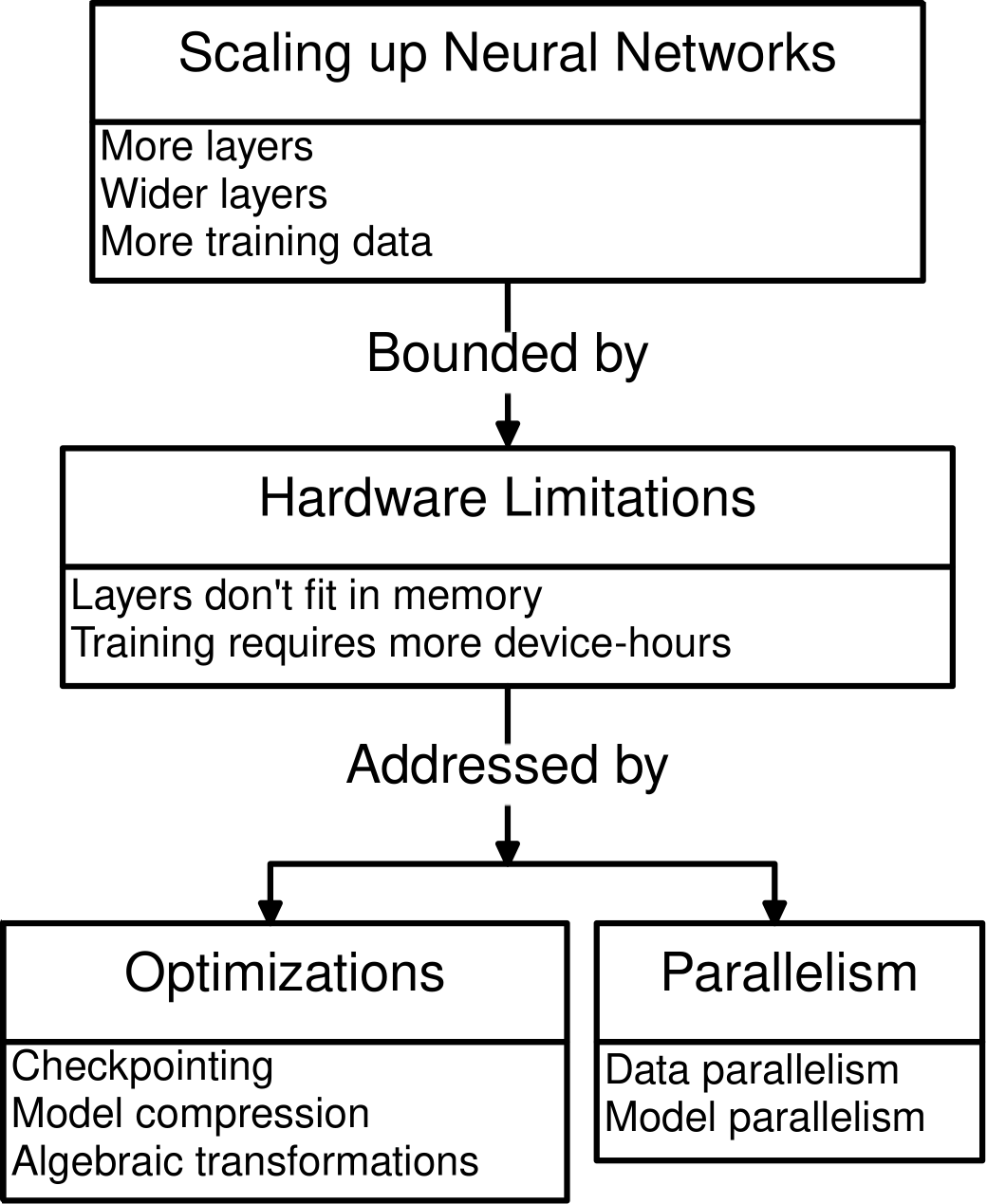}
	\caption{Overview of scaling up NNs within the NN compute infrastructure.}
	\label{fig:nn_scale_up_overview}
\end{figure}

As the field of machine learning has progressed models have become ever larger~\cite{Sevilla:2022:CTAT} and it is through this lens that designing NNs presents an engineering challenge as scaling up a NN has the following effects:
\begin{enumerate}
    \item A larger NN has more neurons performing operations and thus requires more compute.
    \item A larger NN has more parameters and thus requires more memory to store these.
    \item Having more training samples requires more passes and thus more compute.
\end{enumerate}
From this, it is clear that hardware limitations pose limitations for scaling up NNs. In fact, model parallelism is actually amongst the methods aimed at achieving continued progress when it comes to scaling up NNs. An overview is depicted in \Cref{fig:nn_scale_up_overview}.

\begin{figure*}[htbp]
	\centering
	\includegraphics[width=0.65\linewidth]{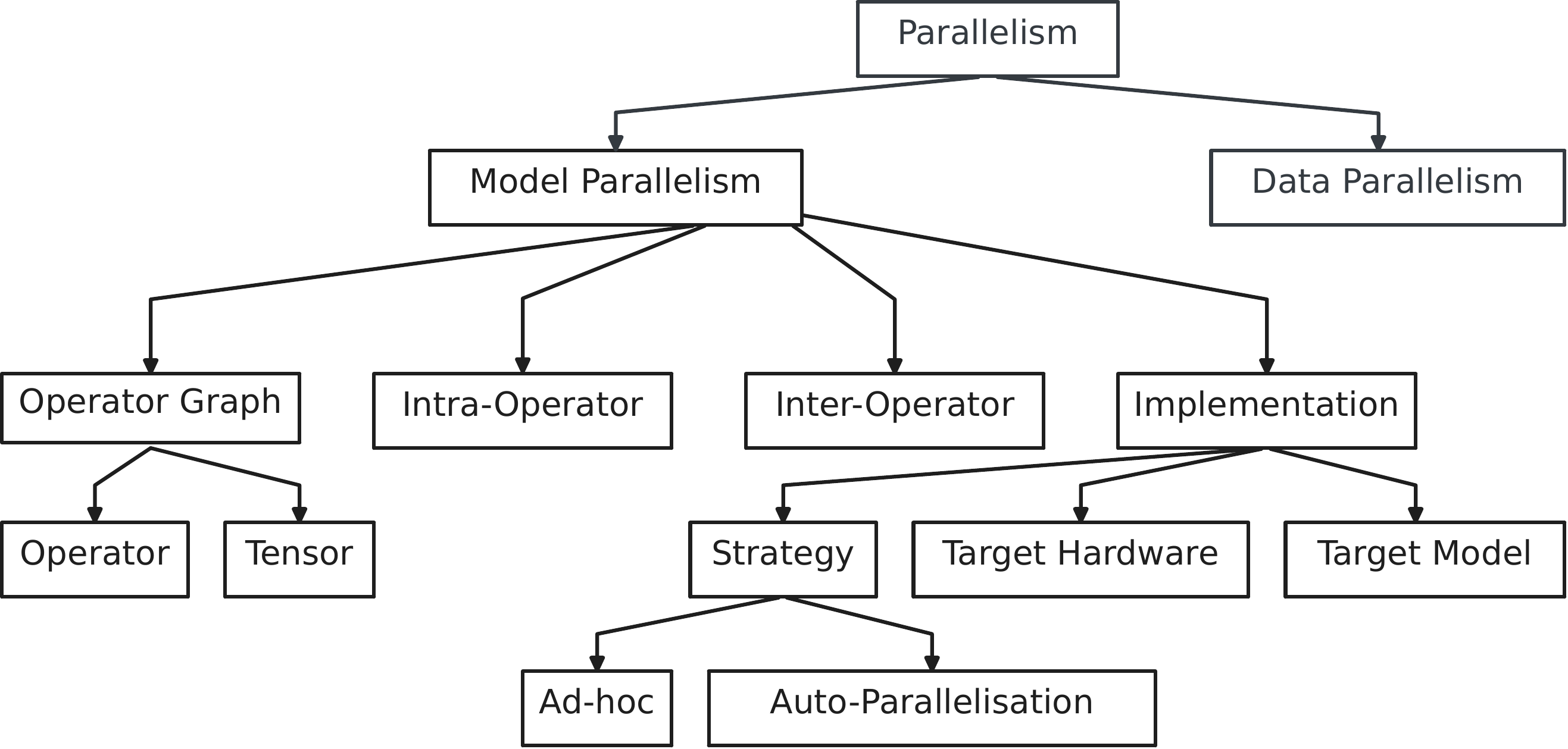}
	\caption{Taxonomy of model parallelism for neural networks. In this survey, we distinguish data parallelism from model parallelism, which both fall under parallelism in neural networks.}
	\label{fig:model_parallelism_taxonomy}
\end{figure*}

We also provide the overview of model parallelism in the form of a taxonomy, \Cref{fig:model_parallelism_taxonomy}, generated from the frequent key-terms.

\subsubsection{Neural network workloads}
\label{subsec:neural_network_workloads}
Compute Flow Graphs (CFGs) are widely used to represent workloads. In keeping with~\cite{Liang:2023:SAPL, Jia:2019:BDMP, Addanki:2019:PLGD}, we will be expressing a NN as a variant of a CFG called an \emph{operator graph}. In this graph, data is represented by tensors and computation by operators.

The term \emph{tensor} is one that comes up a lot in the context of machine learning. While in mathematics a tensor has a more rigorous definition, in the context of NNs, tensors describe n-dimensional arrays of data, flowing through the network. We will regularly mention two named tensors: the input tensor $\mathbf{X}$ and the output tensor $\mathbf{Y}$. Additionally, we distinguish between parameter tensors and activation tensors. Parameter tensors are static inputs to operators, while activation tensors are the result of said operators. Operators are the functional units of the NN. These represent a computation, e.g., a matrix-multiplication or a convolution, performed on any number of input tensors and resulting in a single output tensor. Operators and tensors are organised into an operator graph.

\begin{figure}[htbp]
	\centering
	\begin{subfigure}{\linewidth}
    	\includegraphics[width=\linewidth]{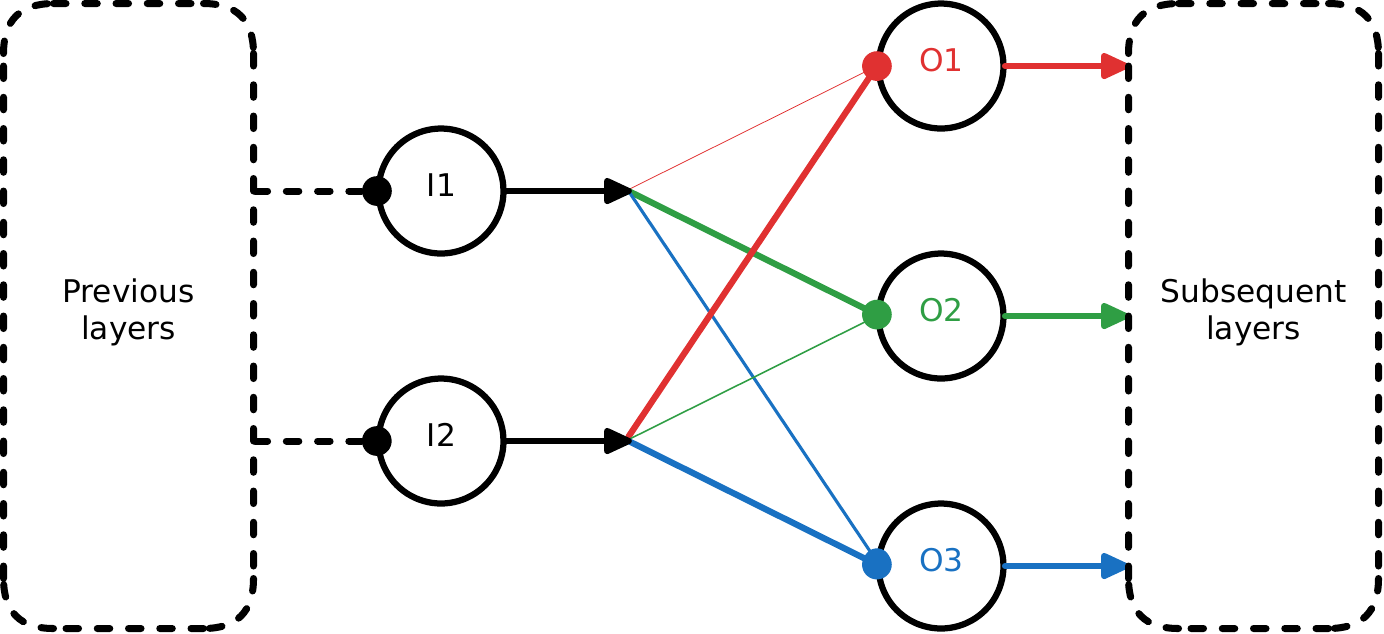}
    	\caption{Neuron representation}
    	\label{fig:fc_neuron_representation}
	\end{subfigure}
	\qquad
	\begin{subfigure}{0.45\linewidth}
		\centering
    	\includegraphics[width=0.7\linewidth]{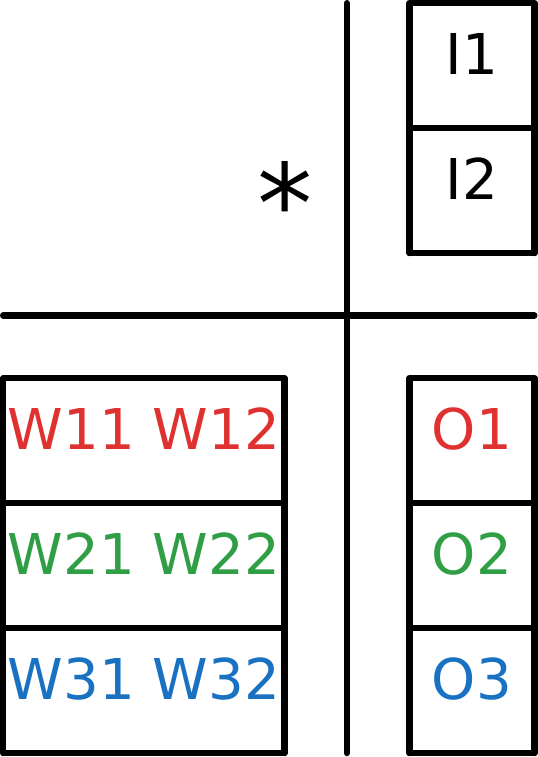}
    	\caption{Tensor representation}
    	\label{fig:fc_tensor_representation}
	\end{subfigure}
	\qquad
	\begin{subfigure}{0.45\linewidth}
		\centering
    	\includegraphics[width=0.7\linewidth]{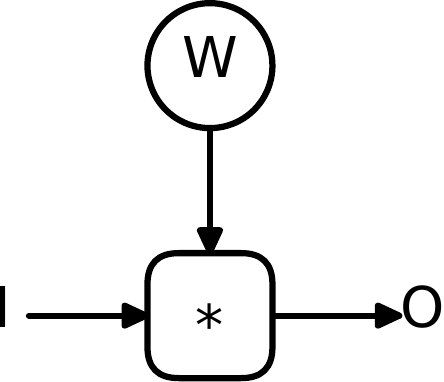}
    	\caption{Operator graph representation}
    	\label{fig:fc_graph_representation}
	\end{subfigure}
	\caption{Three representations of a fully connected layer. The schematic representation highlights the connections between the neurons, the tensor representation shows the mathematical operation implementing the layer, and the operator graph shows the data-flow through the network.}
	\label{fig:fc_layer_three_representations}
\end{figure}

In an operator graph $\mathcal{O} = (V, E)$ of a given NN, every node $v_i \in V$ is either an operator $o_i$, with an associated activation tensor $\mathbf{T}_{o_i}$, or a tensor $\mathbf{T}_i$. Every edge $e_{ij}(v_i, o_j) \in E$ indicates the tensor associated with $v_i$, is an input to the operator node $o_j$. Consider \Cref{fig:fc_layer_three_representations} as a visual explanation. A fully connected layer in a NN (\Cref{fig:fc_neuron_representation}) can be represented as an operation on two tensors, $\mathbf{I}$ and $\mathbf{W}$, resulting in a third tensor, $\mathbf{O}$ (\Cref{fig:fc_tensor_representation}). \Cref{fig:fc_graph_representation} shows how we represent operations on tensors as a graph.

Using this representation we define two workloads: the forward pass and the backward pass. The forward pass takes the input tensor $\mathbf{X}$ and computes all the activations, resulting in $\mathbf{Y}$. The backward pass updates parameters tensors using the back-propagation algorithm. The exact workings for the back-propagation algorithm are beyond the scope of this survey, but we will note the following relevant characteristics:
\begin{itemize}
    \item Starts at the output $\mathbf{Y}$ and works its way back to $\mathbf{X}$, making it dependent on the result of the forward pass.
    \item It requires the activation tensor of every operator to calculate how it should update its parameters.
\end{itemize}
Training consists of $b$ forward passes followed by $b$ backward passes, with $b$ as batch size. Inference only consists of forward passes. 

\subsubsection{Pushing the limits of hardware}
Neural networks have reached a scale, both in terms of compute and memory requirements, where we are arriving at the limits of what current available hardware are capable of. Consequently, there has been significant effort channelled into finding ways to push these limits. Checkpointing~\cite{Chen:2016:TDNS} provides a memory-compute trade-off for the training process. Recall that during the backward pass we require the activation tensor $\mathbf{T}_{o_i}$ for every operator, which were computed during the forward pass. Checkpointing trades some of the memory requirements for storing this for compute by strategically storing only some of the tensors and recomputing the rest from these \emph{checkpointed} tensors. Algebraic transformations~\cite{Unger:2022:UADT} create an equivalent neural network by merging and reordering operators, aiming at a reduction of both computational complexity and memory footprint. 

Both checkpointing and algebraic transformations fully preserve the neural network, but it can also be beneficial to trade some accuracy in representing the network in memory, in order to fit a larger one. For example, using a lower precision data type, such as a 16-bit float instead of a 32-bit float, hurts accuracy. However, the memory saved by this change can be used to store more parameters, which can in turn lead to a greater accuracy gains.

Another approach is to compress a large model into a smaller one~\cite{Rae:2022:SLMM}. This method still requires the training of the larger model variant and thus, contributes only to the inference speed. Pruning is the process of removing unimportant neurons resulting in a sparse network. Which neurons to remove while maintaining model accuracy and how to effectively compute sparse neural network workloads is an active area of research. Distillation is another model compression technique where we try to use a large network in order to train a smaller one, i.e., \emph{distil} the knowledge present in the larger network. Note that while these methods do affect each other, performing algebraic optimisations can potentially hurt parallelism~\cite{Unger:2022:UADT}, as these are not mutually exclusive. In fact, these methods can also complement each other and most works listed in this study do not just utilise model parallelism, but attempt to combine it with other techniques~\cite{Shoeybi:2020:MLMT, Narayanan:2021:ELSL, Smith:2022:UDSM, Rae:2022:SLMM, Chowdhery:2023:PALM, Unger:2022:UADT, Jia:2019:BDMP}. 

A neural network can contain a number of dimensions along which it can be parallelised. For instance, the convolution operator in a CNN often has a number channels which can all be processed in parallel. A model parallelisation strategy then is a mapping from an operator graph to a certain target distributed device (ideally) taking advantage of parallelisable dimensions. Note that due to their parallel nature, it is possible to assign multiple devices to computing a single operator $o_i$. Accordingly, model parallelism encompasses the strategies that utilise parallelisable dimensions within $\mathcal{O}$, while data parallelism are those strategies that utilise parallelisable dimensions in the data. Exactly which parallelisable dimensions are present in any given $\mathcal{O}$ varies greatly and discovering them is a major focus of model parallelism research.

\subsection{Types of model parallelism}
\label{subsec:types_of_model_parallelism}
We present two ways to categorise model parallelism: by the parallelism being exploited with the choice of parallelisation strategy and by the approach employed for finding a specific strategy. In this regard, one can either parallelise over multiple nodes in $\mathcal{O}$, known as \emph{inter-operator parallelism}, or parallelise the operation within an operator node $o_i$, known as \emph{intra-operator parallelism}~\cite{Liang:2023:SAPL}.

\begin{figure}[htbp]
    \centering
    \begin{subfigure}{0.45\linewidth}
    	\centering
        \includegraphics[width=0.55\linewidth]{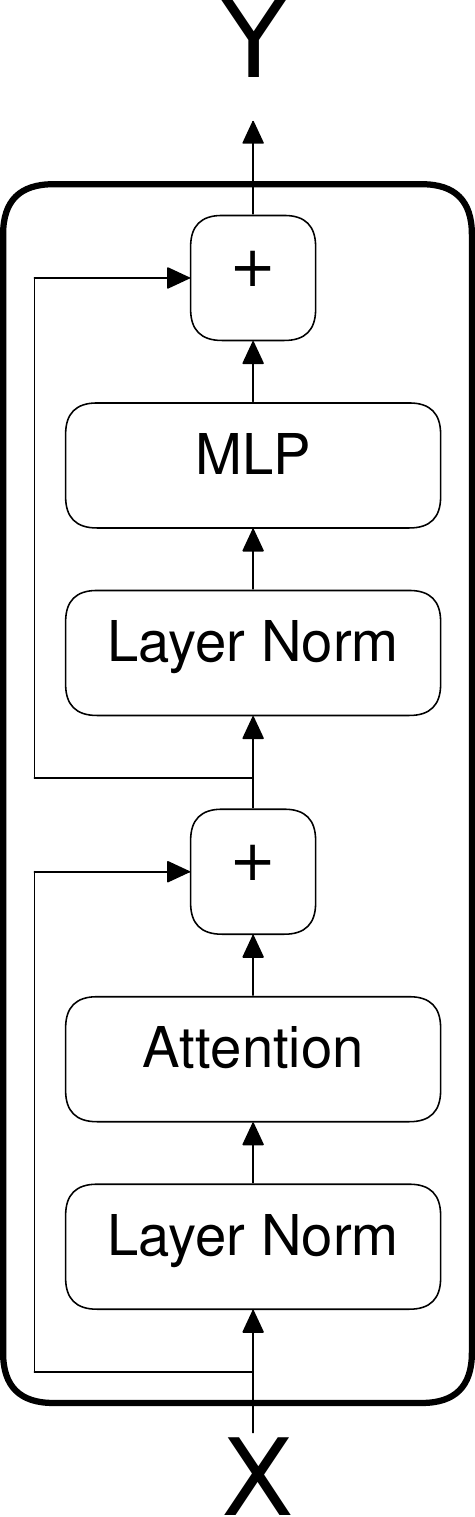}
        \caption{Operator graph of a Transformer layer.}
        \label{fig:operator_graph_transformer}
    \end{subfigure}
    \qquad
    \begin{subfigure}{0.45\linewidth}
    	\centering
        \includegraphics[width=0.65\linewidth]{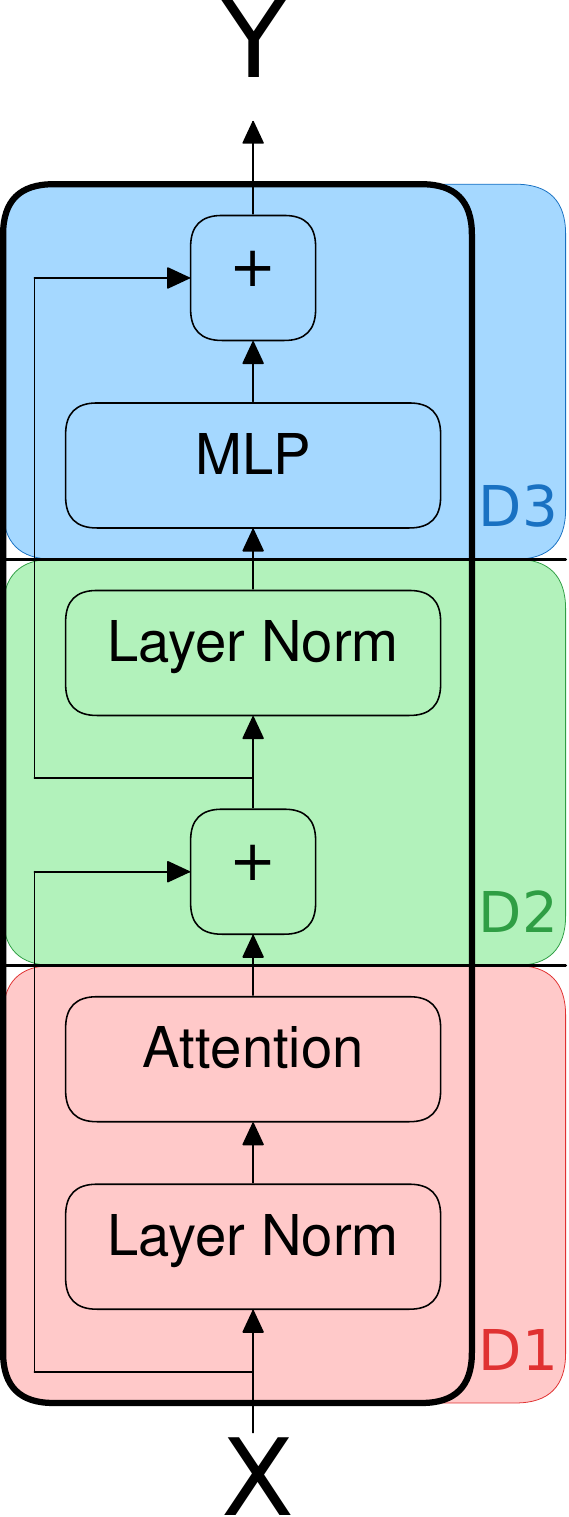}
        \caption{Inter-operator parallelisation of the Transformer layer. The Transformer layer is partitioned over three devices D1, D2 and D3.}
        \label{fig:parallelise_transformer_layer}
    \end{subfigure}
    \qquad
    \begin{subfigure}{\linewidth}
    	\centering
        \includegraphics[width=0.55\linewidth]{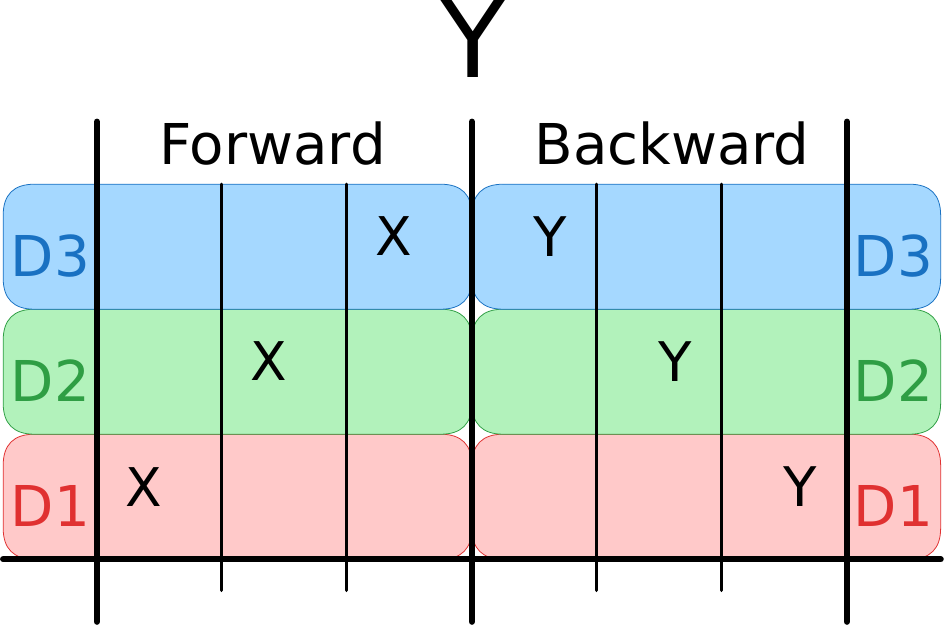}
        \caption{Pipeline view of the inter-operator parallelisation scheme, time progresses from left to right. The input tensor X flows through the partitions on the devices (D1-D3) during the forward pass. The backward pass goes in the opposite direction and depends on the forward pass for its input.}
        \label{fig:pipeline_view_transformer}
    \end{subfigure}
    \qquad
    \begin{subfigure}{\linewidth}
    	\centering
        \includegraphics[width=0.55\linewidth]{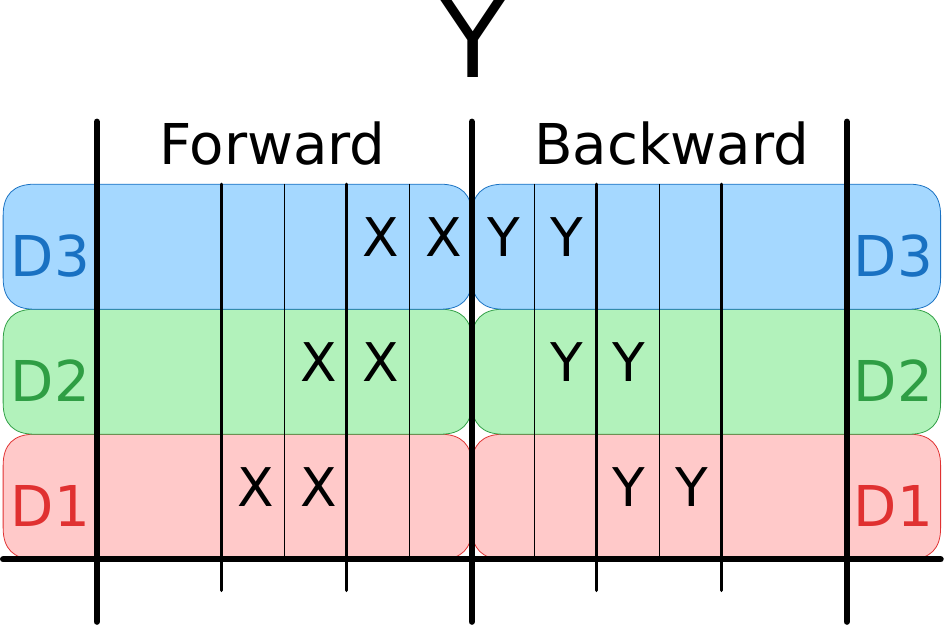}
        \caption{Micro-batches decrease the size of the pipeline bubble. A micro-batch can be sent to the next partition earlier than a full batch, allowing the pipeline to fill up faster, reducing the size of the bubble.}
        \label{fig:micro_batch_transformer}
    \end{subfigure}        
    \caption{Example of a possible inter-operator parallelisation strategy for a Transformer layer and the way an activation tensor flows through it.}
    \label{fig:inter_operator_transformer}
\end{figure}

Inter-operator parallelism essentially comes down to partitioning $\mathcal{O}$ into sub-graphs and assigning every sub-graph to a device. This technique has relatively low communication requirements as we only need to communicate with any other device at the edge of the sub-graph. The parallelisation strategies found in intra-operator parallelism are highly specific to the operator. Again, these two approaches are not mutually exclusive and often are combined into what some call hybrid-parallelism~\cite{Liang:2023:SAPL}. \cite{Shoeybi:2020:MLMT} for example comes up with an intra-operator parallelisation strategy, specifically designed for parallelising a Transformer block. We will explore the reasons behind this approach when we elaborate the challenges of model parallelism in \Cref{sec:challenges}. \Cref{fig:inter_operator_transformer} depicts how an inter-operator strategy would look like when applied to a Transformer layer

Whether to use inter-operator, intra-operator, or a combination of the two, and how exactly to partition a given model using these techniques, depends on many factors, e.g., model architecture and device network topology. Finding the right combination is a major focus of many papers and the approaches taken to achieve an effective strategy is another aspect in which we categorise model parallelism. On the one hand we find \emph{ad hoc approaches} that are specific to a certain model and/or device. On the other hand, we find \emph{general approaches} that work over a variety of models and devices. Ad hoc approaches make use of the target hardware and model architecture being known a priori. An example specifically targeting both Transformer architectures and a hardware with eight A100 GPUs, connected by NVLink, is~\cite{Shoeybi:2020:MLMT}. Another set of examples are papers sponsored by Google, all being specifically designed for Google's TPU pods~\cite{Rae:2022:SLMM, Chowdhery:2023:PALM}. In this context, we notice intra-operator and inter-operator parallelism with a slightly different terminology, i.e., tensor parallelism and pipeline parallelism, respectively.

As mentioned, there are approaches that try to generalise the problem and provide methods for coming up with a strategy for any $\mathcal{O}$ on any distributed device. Some methods require the user to specify the strategy~\cite{Huang:2019:GPET, Shazeer:2018:MTDL}, while others are fully automated~\cite{Jia:2019:BDMP}.

\section{Challenges}
\label{sec:challenges}
Considering what stated so far and based on our covered literature, major challenges affecting auto-parallelisation are listed below.

\paragraph{Inter-operator parallelism}
Inter-operator parallelism suffers from low device utilisation if the implementation does not make use of pipelining. After all, the input of each partition is the output of a previous one and processing can only start once this previous partition has produced said output. In addition to the complexities imposed by a pipeline, which is beyond the scope of this survey, pipelines encounter frequent stalls, i.e., \emph{bubbles}, during training~\cite{Huang:2019:GPET} (\Cref{fig:pipeline_view_transformer}). Due to the data-dependency between backward and forward passes, the former can't be started until the latter is completed.

\paragraph{Intra-operator parallelism}
Intra-operator parallelism's challenge is in its extreme communication requirements. The input tensor to the parallelised operator needs to be scattered over the devices and the output then needs to be gathered for every batch. 

\paragraph{Combining parallelism types}
As it is concluded in~\cite{Smith:2022:UDSM}, every form of parallelism, including data parallelism, has its own limitations and many implementations end up using hybrid strategies. Such a strategy can be seen in \Cref{tab:transformer_results,tab:auto_parallelisation_frameworks}. Jia et al.~\cite{Jia:2019:BDMP} note that available deep learning frameworks are often simple and suboptimal when it comes to parallelising models. This makes exploring different hybrid strategies a significant challenge. General parallelism approaches attempt to solve this either in the form of fully automated auto-parallelisation frameworks (\Cref{tab:auto_parallelisation_frameworks}), or by providing a more high-level programming model, simplifying the expression of the intended parallelisation strategy~\cite{Xu:2021:GSPM, Huang:2019:GPET}.

Auto-parallelisation generally is expressed as a search problem, bringing along the usual challenges attached to search problems. We will now briefly list these as they pertain to model parallelism.

\paragraph{Search-space}
The search space in DNN auto-parallelisation is the set of strategies that can be evaluated. A good definition allows for strategies to exploit a large amount of parallelisable dimensions, while excluding illegal and/or suboptimal strategies.

\paragraph{Strategy evaluation}
Given the need to quickly traverse the search-space, fully profiling every parallelisation strategy is not computationally feasible, which is why the performance of the strategy must be \emph{estimated} in some way. While compute and memory are relatively easy to predict~\cite{Jia:2018:EHDA, Jia:2019:BDMP}, modelling communication time based on the network latency and bandwidth of whatever cluster medium is being used, is currently a major open challenge.

\paragraph{Search method}
Finding optimal methods for traversing the search-space is a challenge in itself and the approaches taken in the context of model parallelism have scattered in many directions, as it can be seen later in \Cref{tab:auto_parallelisation_frameworks}.

\section{Use-cases}
\label{sec:use_cases}
It is generally known that a model's accuracy improves as it get bigger and trains over more data. Interestingly, it is shown that large-scale Transformers for natural language processing tasks, colloquially known as Large Language Models (LLMs), show exceptional performance in few-shot learning applications~\cite{Brown:2020:LMFS}. Since the release of GPT-3, followed by the availability of GPT-3.5 and GPT-4 to masses in the form of ChatGPT, the technology industry has, at the time of this writing, seen a renewed effort to scale up models. This makes LLMs a prominent use-case for model parallelism as these models have now scaled well beyond the capabilities of a single device, both in terms of memory and compute. The details of our selected use-case models are listed in \Cref{tab:transformer_results}.
\begin{table*}[htbp]
	\centering
	\caption{Model size, parallelism type and hardware utilisation achieved for ad hoc approaches when scaling up Transformer models. Though~\cite{Anil:2023:PALM, Brown:2020:LMFS, OpenAI:2023:GPT4} do not provide implementation detail of their model architectures, they are included for completeness.}
	\begin{tabular}{l|lrlrrrr}
	\toprule
	 					& & & & \multicolumn{3}{c}{\textbf{Parallelism}} \\
 	\cmidrule(lr){5-7}
	\multicolumn{1}{c|}{\textbf{Model family}} & 
	\multicolumn{1}{c}{\textbf{Paper}} & 
	\multicolumn{1}{c}{\textbf{\begin{tabular}[c]{@{}c@{}}Largest model\\ (\# parameters)\end{tabular}}} & 
	\multicolumn{1}{c}{\textbf{Training hardware}} & 
	\multicolumn{1}{c}{\textbf{\begin{tabular}[c]{@{}c@{}}Intra-\\ operator\end{tabular}}} & 
	\multicolumn{1}{c}{\textbf{\begin{tabular}[c]{@{}c@{}}Inter-\\ operator\end{tabular}}} & 
	\multicolumn{1}{c}{\textbf{Data}} &
	\multicolumn{1}{c}{\textbf{Utilisation}} \\
	\midrule
	\multirow{4}{*}{Megatron} 		& \cite{Shoeybi:2020:MLMT} 		& 8.3B 		& 32$\times$16 V100s 		& 8 		& 1 	& 64 				& <30\% (hardware) \\
                          			& \cite{Narayanan:2021:ELSL} 		& 1T 		& 8$\times$384 A100s 		& 8 		& 64 	& 6 (presumed) 		& 52\% (hardware) \\
                          			& \cite{Smith:2022:UDSM}         	& 530B 		& 8$\times$420 A100s 		& 8 		& 35 	& 12 				& 36.2\% (hardware) \\ 
                          			& \cite{Korthikanti:2023:RARL} 	& 1T 		& 8$\times$64 A100s 		& 8 		& 64 	& 1 				& 56.3\% (model) \\
	\midrule
	\multirow{3}{*}{Gopher/PaLM}	& \cite{Rae:2022:SLMM} 			& 280B 		& 4$\times$1024 TPUv3s 	& >1 		& 4 	& >1 				& n/a \\
	     							& \cite{Chowdhery:2023:PALM} 		& 540B 		& 2$\times$3072 TPUv4s 	& 12 		& 1 	& 2$\times$256 		& 46.2\% (model) \\
                          			& \cite{Anil:2023:PALM} 			& <540B  	& n/a TPUv4s 				& n/a 		& n/a 	& n/a 				& n/a \\
	\midrule
	\multirow{2}{*}{GPT}      		& \cite{Brown:2020:LMFS} 			& 175B 		& n/a V100s 				& n/a 		& n/a 	& n/a 				& n/a \\
                          			& \cite{OpenAI:2023:GPT4} 			& n/a 		& n/a 						& n/a 		& n/a 	& n/a 				& n/a \\
	\bottomrule
	\end{tabular}
	\label{tab:transformer_results}
\end{table*}

\begin{figure}[htbp]
	\centering
	\begin{subfigure}{\linewidth}
	    \centering
	    \includegraphics[width=0.6\linewidth]{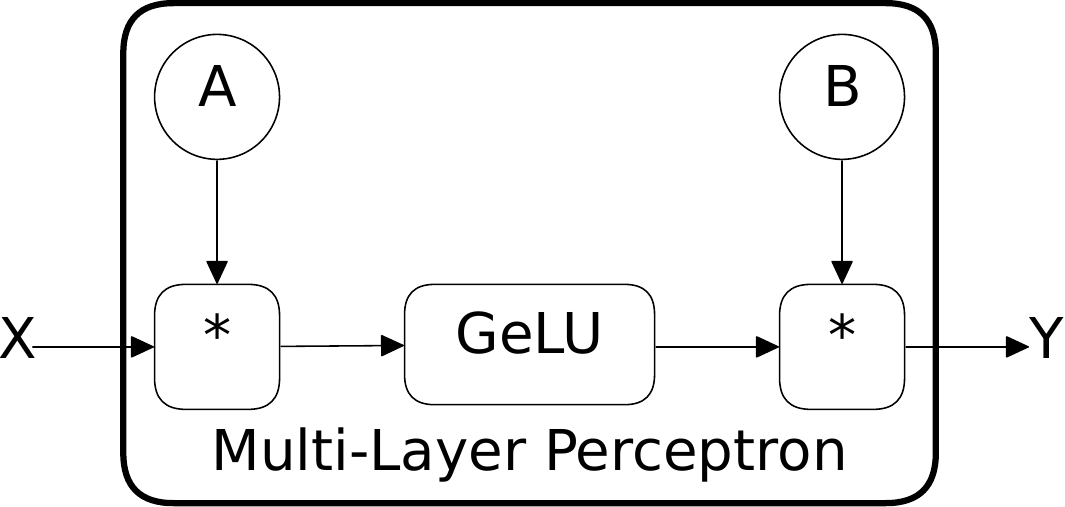}
	    \caption{Graphical representation of the MLP operator graph.}
	    \label{fig:mlp_operator_graph}
	\end{subfigure}
	\qquad
	\begin{subfigure}{\linewidth}
	    \centering
	    \includegraphics[width=0.6\linewidth]{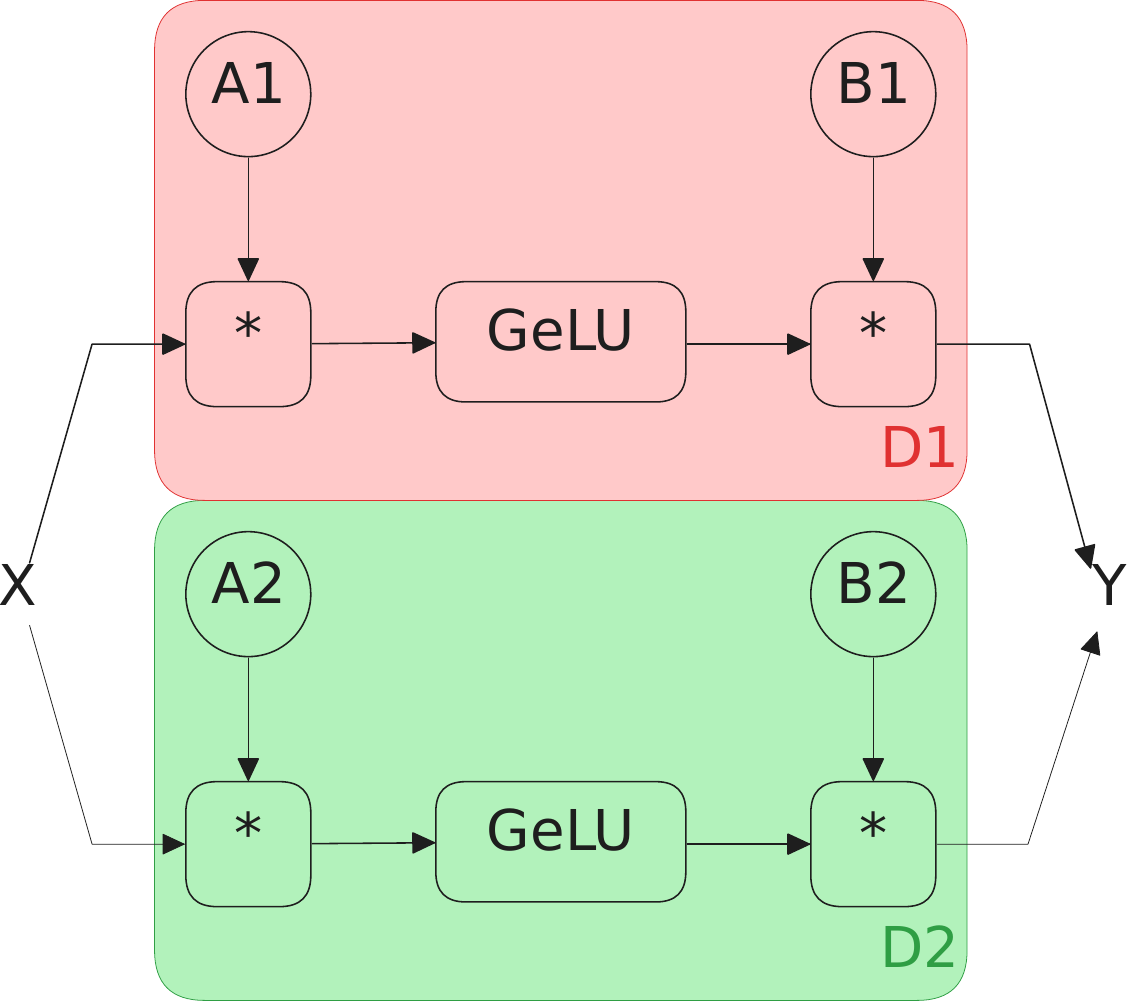}
	    \caption{Intra-Operator parallelisation by splitting $A$ along the columns. In this case, no communication is needed within the operator.}
	    \label{fig:mlp_intra_operator_good}
	\end{subfigure}
	\qquad
	\begin{subfigure}{\linewidth}
	    \centering
	    \includegraphics[width=0.7\linewidth]{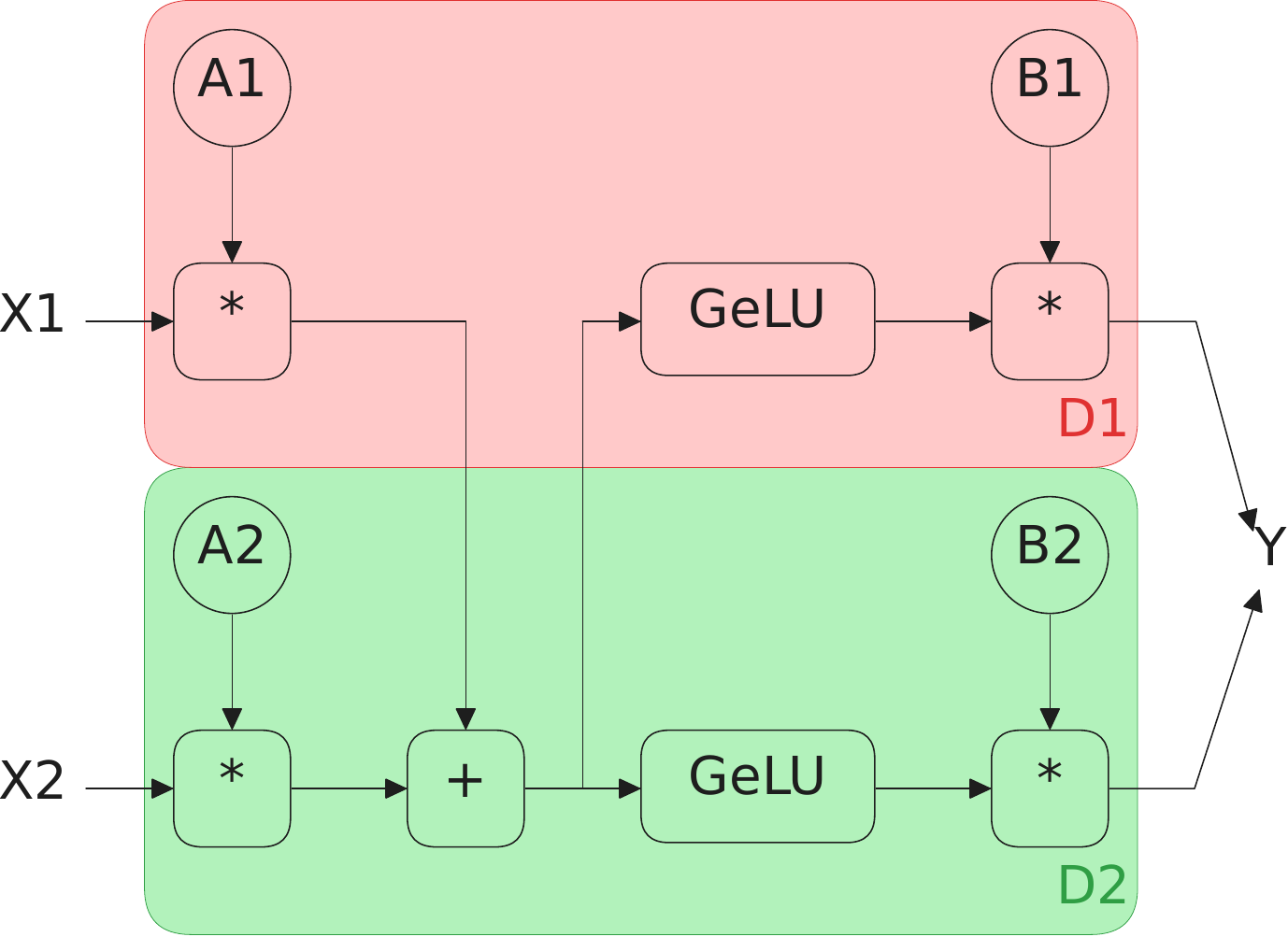}
	    \caption{Intra-Operator parallelisation by splitting $A$ along the rows. Since the \texttt{GeLU} operator is non-linear, the activation tensors have to be gathered and scattered first.}
	    \label{fig:mlp_intra_operator_bad}
	\end{subfigure}
	\caption{The operator graph of the Multi-Layer Perceptron (MLP) and the two intra-operator strategies for it~\cite{Shoeybi:2020:MLMT}.}
	\label{fig:mlp_operator_graphs}
\end{figure}

Methods in this section are all expert designs and highly specific to the Transformer architecture. As a case-study however, these do provide valuable insights into the challenges of model parallelism. First, we will recap an important building block of neural networks, i.e., Multi-Layer Perceptron (MLP). The MLP consists of four operators. A fully connected layer, followed by a \texttt{GeLU()} activation function, followed by another fully connected layer, followed by a \texttt{Dropout()} function. The \texttt{Dropout()} is only used during training and we will skip it here. More formally, we could note an MLP as,
\begin{align*}
	Z &= MLP_{A,B}(X) \text{,} \\
	&= \texttt{GeLU}(X \cdot A)\cdot B \text{.}
\end{align*}
Where $A$ and $B$ are the weight matrices of the fully connected layers. This is represented visually as an operator graph in \Cref{fig:mlp_operator_graph}, alongside different intra-operator strategies applied to MLPs \Cref{fig:mlp_operator_graphs}.

\subsection{Megatron}
Shoeybi et al.~\cite{Shoeybi:2020:MLMT} present a technique to partition large Transformer models over multiple GPUs. They demonstrate their approach by training two 8.3B parameter (GPT-2) and  3.9B parameter (BERT) models on up to eight GPUs. This was implemented using a form of intra-layer parallelism where the two building blocks of the Transformer model, the MLP and the self-attention, are distributed over multiple GPUs. 

Two approaches are considered for the MLP, splitting the weights over the columns or over the rows. These approaches are visually depicted in \Cref{fig:mlp_intra_operator_good,fig:mlp_intra_operator_bad}. If we were to distribute $A$ over the rows, we get
\begin{align*}
	X &= [X_1, X_2] \text{ and} \\
	A &= \begin{bmatrix} A_1\\ A_2 \end{bmatrix} \text{.}
\end{align*}
This would mean that we calculate \texttt{GeLU()} as,
\[
Y = \texttt{GeLU}(X_{1} A_{1} + X_{2} A_{2}) \text{,}
\]
and because \texttt{GeLU()} by design is non-linear,
\[
\texttt{GeLU}(X_{1} A_{1} + X_{2} A_{2}) \neq \texttt{GeLU}(X_{1} A_{1}) + \texttt{GeLU}(X_{2} A_{2}) \text{,}
\]
which means $X_{1} A_{1} + X_{2} A_{2}$ needs to be calculated before we are able to calculate \texttt{GeLU()}. Accordingly, such a calculation requires the sending of either $X_{1} A_{1}$ or $X_{2} A_{2}$ over the network.

Conversely distributing $A$ over the columns as,
\[
A = [A_1, A_2] \text{,}
\]
allows the calculation of \texttt{GeLU()} as,
\[
\texttt{GeLU}(X A) = [\texttt{GeLU}(X A_1), \texttt{GeLU}(X A_2)] \text{.}
\]
The next step is to distribute $B$ over the rows as, 
\[
Z = \texttt{Dropout}(\texttt{GeLU}(X A_1) \cdot B_1 + \texttt{GeLU}(X A_2) \cdot B_2) \text{,}
\]
eliminating the need to reconstruct $\texttt{GeLU}(X A)$ altogether. Naturally, this was the approach that the authors have opted to use.

Since the self-attention block of the Transformer has a lot of inherent parallelism, the $Q$, $K$, and $V$ matrices are simply distributed over the columns. To demonstrate their approach, the authors have trained an 8.9B GPT-2 Transformer model on eight GPUs with 77\% performance scaling in terms of throughput.

Narayanan et al.~\cite{Narayanan:2021:ELSL}, build on the 8-way intra-layer parallelism technique from~\cite{Shoeybi:2020:MLMT} and combine it with up to 64-way inter-layer parallelism using an approach similar to~\cite{Narayanan:2019:PDGP}, in order to fit up to a 1T parameter model on A100 GPUs. With the addition of data parallelism, the authors manage to achieve 52\% hardware utilisation for the largest model. Additionally, they analyse combining parallelism techniques using an analytical model and provide three key takeaways:
\begin{itemize}
    \item When considering different forms of model parallelism, tensor (intra-layer) model parallelism should generally be used up to degree $g$ when using $g$-GPU servers, and then pipeline (inter-layer) model parallelism can be used to scale up larger models across servers.
    \item The optimal microbatch size $b$, depends on the throughput and memory footprint characteristics of the model, as well as the pipeline depth $p$, data-parallel size $d$, and batch size $B$.
    \item When using data and model parallelism, a total model-parallel size of $M = t \cdot p$ should be used so that the model's parameters and intermediate metadata fit in GPU memory. Data parallelism can be used to scale up training to more GPUs.
\end{itemize}

Smith et al.'s paper~\cite{Smith:2022:UDSM} is a combined effort from NVIDIA and Microsoft to train a large language model by combining the former's Megatron framework with the latter's DeepSpeed framework. The authors utilise data, inter-layer, and intra-layer parallelisms to train up to a 540B parameter model on 420 DGX A100 servers containing eight A100s each.

We also have the research from Korthikanti et al.~\cite{Korthikanti:2023:RARL}, which details fitting a 1T parameter model. The main focus of this paper is reducing activation memory and increasing parallelism through two new parallelisation schemes. The activation memory is defined as the memory that is required to store the tensor created during the forward pass of the training algorithm. This does not include the model parameters. For an input tensor $X \in \mathbb{R}^{s \cdot b \cdot h}$, where $s$ is the sequence length, $b$ is the micro-batch size, $h$ is the hidden dimension size and $a$ being the number of attention heads, the attention block has an activation memory footprint of $11 s \cdot b \cdot h + 5 a \cdot s^2 \cdot b$ Bytes. The MLP block and the layer-norm have footprints of $19 s \cdot b \cdot h$ and $4 s \cdot b \cdot h$ Bytes, respectively. Accordingly, the total activation memory footprint for a single layer in a Transformer model then is
\[
s \cdot b \cdot h(34 + 5 \frac{a \cdot s}{h}) \text{.}
\]

The tensor parallelism from the previous approach is used again as it is computationally efficient. Parts of the layer that are computationally most expensive are parallelised. It also parallelises the activations within these blocks, meaning that the per device memory footprint can be expressed as,
\[
s \cdot b \cdot h(10 + \frac{24}{t} + 5\frac{a \cdot s}{ht}) \text{.}
\]
It does not however parallelise the \texttt{Dropout} or layer-norms. Hence, even when $\lim_{t \to \infty}$, the activation memory footprint is still $10 s \cdot b \cdot h$. The researchers note that the \texttt{Dropout} and layer-norm operations are independent along the sequence dimension and partition the activation tensor accordingly. They name this approach \emph{sequence parallelism}. Now, the per-device memory footprint of a layer is expressed by,
\[
\frac{s \cdot b \cdot h}{t}(34 + 5\frac{a \cdot s}{h}) \text{.}
\]

Lastly, pipeline parallelism is introduced. The concept of pipeline parallelism has been discussed before, however, it does have an implication on the activation memory footprint. For the first-stage, the memory footprint is,
\[
\frac{s \cdot b \cdot h \cdot L}{t}(34 + 5\frac{a \cdot s}{h}) \text{,}
\]
where $L$ is the number of layers in the network. For subsequent stages, memory requirements are slightly different. Although not the entire footprint is captured by this equation, it does so for the overwhelming majority and for simplicity sake, the authors use this equation to reason about their implementation. To show the effectiveness of their approach, four Transformer models are trained ranging from 22B to 1T parameters on 8 and 512 GPUs, respectively. The authors manage to achieve 41.5\% hardware utilisation for the smallest model and 56.3\% for the largest model, without the use of any data parallelism.

\subsection{Gopher}
Rae et al.'s work~\cite{Rae:2022:SLMM} is an effort by Google to train a large language model. Their approach differs in that the hardware and software are custom. The authors use TPU hardware and custom JAX software framework. The largest Gopher model uses 4-way inter-layer parallelism over four TPU pods, as well as an unreported level of intra-layer parallelism and data parallelism within a 1024 chip pod, in order to train a model with up to 280B parameters.

\subsection{PaLM}
Chowdhery et al.'s work~\cite{Chowdhery:2023:PALM} is another effort by Google to train a large language model on Google TPUs. The authors train a model with up to 540B parameters using two TPUv4 pods, consisting of 3072 chips each. Notable here is that the authors do not use any inter-layer parallelism, thereby avoiding the pipeline bubble problem during training. They use up to 12-way intra-layer parallelism and 256-way data parallelism within a single pod and another 2-way data parallelism to scale up to two pods. In terms of software, the authors use their own Pathways framework~\cite{Barham:2022:PADD}, which is built on top of JAX.

The paper~\cite{Anil:2023:PALM} details the next version of the PaLM model, i.e., PaLM 2. Unfortunately Google has opted not to share any detail about this model's underlying compute infrastructure. We only include it here for completeness sake.

\subsection{GPT}
While model parallelism is certainly employed by OpenAI for their GPT-3 model, for both training and inference, the employed V100 GPUs~\cite{Brown:2020:LMFS} lack the memory to store the model in its entirety. We can simply list these here for completeness sake, as OpenAI has opted not to share any detail about their training infrastructure. Similarly, no detail about the next-generation GPT-4~\cite{OpenAI:2023:GPT4} model, nor the infrastructure behind the model have been released.

\section{Discussion}
\label{sec:discussion}
As mentioned in \Cref{sec:challenges}, the communication costs of intra-operator parallelism is so high that it is only possible to achieve it with the use of high-speed interconnects. Within the Megatron family, the same 8-way intra-operator parallelism for Transformer layers by~\cite{Narayanan:2021:ELSL} is used, a deep dive of which is provided in \Cref{subsec:types_of_model_parallelism}. This approach relies on NVLink interconnects between the devices, limiting it to a single compute node. For the Gopher/PaLM family however, different hardware is employed, making it possible to apply up to 12-way intra-operator parallelism~\cite{Chowdhery:2023:PALM}. These clusters are specifically designed for neural network workloads and have very high speed interconnects. 

As discussed in \Cref{sec:challenges}, \cite{Narayanan:2019:PDGP} mitigates the pipeline bubble by keeping multiple batches in-flight and scheduling them asynchronously. Alternatively, \cite{Huang:2019:GPET} takes a different approach, noting that there is parallelism within a batch. The authors subdivide batches into micro-batches, which can be pipelined much more efficiently, as depicted in \Cref{fig:micro_batch_transformer}. 

Considering \Cref{tab:transformer_results}, while implementations within a family share details such as hardware and model architecture, different ad hoc approaches have very few similarities, making them hard to compare. To address this limitation, the use of the Model FLOPs Utilisation (MFU) metric over Hardware FLOPs Utilisation (HFU) is proposed in~\cite{Chowdhery:2023:PALM}. This metric takes into account that frequently employed techniques such as remetarialisation are used to trade off memory usage with compute. This creates a scenario where using additional hardware FLOPs can save memory, increasing HFU, without having an actual impact on the overall throughput of the system. MFU is based on the actual throughput of the system (tokens per second in the case of Transformers) compared to the theoretical maximum of the system. This has been picked up on by the latest Megatron paper~\cite{Korthikanti:2023:RARL}. However, Google's authors in their next paper~\cite{Anil:2023:PALM}, do not report anything about hardware configuration, let alone the MFU they are able to achieve.

As is noted in~\cite{Smith:2022:UDSM}, none of the three forms of parallelism in neural networks can address all the challenges in training billion parameter models and indeed, we see in \Cref{tab:transformer_results} that of the papers in the Megatron family, at least two forms of parallelism is considered. The two papers that use the most GPUs by far, both employ all three forms. Similarly, while considering their proprietary TPU hardware, Google manages to avoid using inter-layer parallelism entirely for PaLM and only uses 4-way for Gopher. They still heavily rely on data parallelism in order to maintain throughput while scaling to thousands of TPUs. 

Zheng et al.~\cite{Zheng:2022:AIIO} provide a comparison between ad hoc and general strategies, comparing the approach with Megatron-LM~\cite{Narayanan:2021:ELSL} and DeepSpeed~\cite{Rasley:2020:DSSO}. Their Alpa is able to match the former and outperform the latter. We must note the absence of any comparison between Alpa, FlexFlow~\cite{Jia:2019:BDMP} and Tofu~\cite{Wang:2019:SVLM}. This is due to the fact that at the time of their publication, FlexFlow did not support the required operators and Tofu had not released their source code. 

Thus, while \Cref{tab:auto_parallelisation_frameworks} offers a comparison of listed frameworks, the lack of standardised testing means that it is very hard to draw any conclusions about how the different approaches actually compare on specific metrics. Search methods especially have scattered into all directions and it is almost impossible to discern which one is better, since the search-space is defined differently for every case. 
\begin{table*}[htbp]
    \centering
    \caption{Overview of parallelisation frameworks, automatic and manual.}
    \begin{tabular}{l|lccll}
    \toprule
    						& & \multicolumn{2}{c}{\textbf{Parallelism}} \\
    \cmidrule(lr){3-4}
    \multicolumn{1}{c|}{\textbf{Mode}}	 & 
    \multicolumn{1}{c}{\textbf{Framework/Paper(s)}} & 
    \multicolumn{1}{c}{\textbf{Intra-operator}} & 
    \multicolumn{1}{c}{\textbf{Inter-operator}} & 
    \multicolumn{1}{c}{\textbf{Search method}} & 
    \multicolumn{1}{c}{\textbf{Strategy evaluation}} \\
    \midrule
    \multirow{6}{*}{Automatic} 	& RaNNC~\cite{Tanaka:2021:AGPV} 							& - 			& \checkmark 	& Dynamic programming 			& Profiling operators \\
    								& FTPipe~\cite{Eliad:2021:FTGN} 							& - 			& \checkmark 	& Multi-processor scheduling 	& Profiling operators \\
    								& Alpa~\cite{Zheng:2022:AIIO} 								& \checkmark 	& \checkmark 	& Dynamic programming 			& Profiling-calibrated model \\
    								& FlexFlow~\cite{Jia:2019:BDMP, Unger:2022:UADT} 			& \checkmark  	& \checkmark 	& Markov chain Monte Carlo 	& Calibrated simulation \\
    								& TensorOpt~\cite{Cai:2022:ETDT} 							& \checkmark 	& - 			& Frontier tracking 			& Profiling-calibrated model \\
    								& Double recursive~\cite{Wang:2021:ESPL} 					& \checkmark 	& - 			& Double recursive 				& Symbolic model \\
    \midrule
    \multirow{3}{*}{Manual}		& GPipe~\cite{Huang:2019:GPET} 							& - 			& \checkmark 	& - 							& - \\
    								& PipeDream~\cite{Harlap:2018:PDFE, Narayanan:2019:PDGP} 	& - 			& \checkmark 	& - 							& - \\
    								& GSPMD~\cite{Xu:2021:GSPM} 								& \checkmark 	& - 			& - 							& - \\
    \bottomrule
    \end{tabular}
    \label{tab:auto_parallelisation_frameworks}
\end{table*}

As discussed, comparing papers from DNN auto-parallelisation quantitatively poses its challenges. Alternatively, we turn to a qualitative analysis of the papers found in this table. Tanaka et al.~\cite{Tanaka:2021:AGPV} employ dynamic programming to automatically partition any model formatted in the PyTorch model specification into a number of subgraphs. These subgraphs are load-balanced under the constraint of the available memory on the devices at hand. Eliad et al.~\cite{Eliad:2021:FTGN} consider automatic inter-layer parallelism to create a framework to fine-tune models for commodity hardware. They extend the strategy-space of PipeDream by allowing non-adjacent layers to be scheduled onto the same GPU. This means that pipeline stages can be made smaller, allowing for more fine-grained load-balancing at the expense of increased communication overhead. The authors use four competing search methods to explore the new strategy-space. Three existing methods are considered (as listed below), as well as one new search algorithm, specifically tailored to their search-space.
\begin{itemize}
    \item PipeDream~\cite{Narayanan:2019:PDGP}: Exhaustive search
    \item Acyclic~\cite{Moreira:2017:GPAC, Moreira:2018:EMLA}: Greedy search
    \item Metis~\cite{Schloegel:2002:PSDM}: General graph partitioning scheme
\end{itemize}
These strategies are evaluated by profiling every operator in the graph in isolation and utilising this data to calibrate a cost model. Compared to PipeDream's partitioning scheme, FTPipe Mixed-pipe is able to fit a 3B parameter model on eight RTX 2080-Ti GPUs with 11GB of memory each, connected over a PCI-e 3.0 bus. The PipeDream partitioning scheme did not yield a valid parallelisation strategy on this setup.

Zheng et al.~\cite{Zheng:2022:AIIO} motivate their hierarchical search-space by noting that \enquote{intra-layer and inter-layer parallelism take place at different granularities of the DL computation and have distinct communication requirements, which happen to match the structure of today's compute clusters}. The structure they refer to is a mesh network. The search space is formulated as a two-level hierarchy in order to express both inter- and intra-layer parallelism strategies. The lowest level of the hierarchy Alpa takes an operator graph, a device mesh and chooses an intra-layer strategy for every node in the graph, such that the total execution cost of the graph is minimised. It formulates this as an integer linear programming problem and consider an off-the-shelf-solver, able to efficiently solve for graphs consisting of thousands of operators. The second level consists of finding an optimal partitioning of the operator graph and mapping this to a sub-mesh of the compute cluster. The search method used here is a dynamic programming algorithm that takes as reward the predicted performance of a stage-mesh pair, optimised by the lower level.

Jia et al.'s work~\cite{Jia:2019:BDMP} details an entirely new framework built from the ground up for auto-parallelisation. This framework, called FlexFlow, introduces itself with a deep learning engine that uses a comprehensive search-space of parallelisation strategies, called \emph{SOAP}. The SOAP search-space consists of four parallelisable dimensions: Sample, Operator, Attribute, and Parameter. We had mentioned the concept of parallelisable dimensions in \Cref{subsec:background}.
\begin{itemize}
    \item \emph{Sample dimension} describes the amount of data samples.
    \item \emph{Parameter dimension} is defined as requiring the splitting of model parameters. We know this as intra-layer parallelism.
    \item \emph{Attribute dimension} does not require the splitting of model parameters. This essentially is a catch-all dimension when there are additional ways to parallelise an operator.
    \item \emph{Operator dimension} which represents operators wholly.
\end{itemize}

As already alluded to by the description of the attribute dimension, not every dimension exists for every operator's output tensor. Some dimensions may have multiple axes along which parallelisation could occur. As such, the full SOAP search-space consists of the set $\mathcal{P}$, comprising of ordered sets of parallelisable dimensions, $\mathcal{P}_i$, which are mapped to the elements of $\mathcal{O}$ in an injective manner. The search-space $\mathcal{P}$ can be formulated as
\[
\mathcal{P} = \{f(o_i) | \forall o_i, o_j \in \mathcal{O}, o_i = o_j \implies f(o_i) = f(o_j)\} \text{.}
\]

Accordingly, a strategy $\mathcal{S}$ in FlexFlow is defined as a set of positive integer tuples, $c_i$, such that
\[
\mathcal{S} = \{c_i | \forall \mathcal{P}_i \in \mathcal{P}, c_i \in \mathbb{Z}^{|\mathcal{P}_i|} \} \text{.}
\]
Here, $c_i$ describes the degree of parallelism for each of the parallelisable dimensions present in $P_i$, resulting in a number of independent tasks equal to the product of the tuple's elements. While other works classify FlexFlow's search-space as containing intra-layer parallelism only~\cite{Liang:2023:SAPL}, arguing it does not support pipeline parallelism, we do include inter-layer parallelism. FlexFlow is capable of organising operators from the operator graph into subgraphs and assigning these to different devices. To evaluate parallelisation strategies found in the search-space, FlexFlow utilises an execution simulator, taking as input,
\begin{itemize}
    \item a device graph,
    \item an operator graph, and
    \item a parallelisation strategy.
\end{itemize} 

The first step is to construct a task graph, $\mathcal{T}$, from the three inputs. In the task graph, nodes represent tasks as defined by the strategy, while edges represent a dependency between two tasks. One important detail to note is that unlike the operator graph, edges here do not represent the flow of data, but just the partial ordering of the task set. The simulator uses a combination of profiling tasks on target devices, estimating communication overhead from the size of the tensors and the characteristics of the device connections. This process provides an estimate for the total execution time of the task graph. As the search method, FlexFlow employs a Markov Chain Monte Carlo search algorithm, i.e., randomly sampling both operators and strategies, followed by evaluation using the simulator described above. FlexFlow is directly compatible with models specified in PyTorch format, but also includes front-ends for both ONNX and TensorFlow Keras support.

Cai et al.~\cite{Cai:2022:ETDT} optimise for both memory consumption and execution time, providing a Pareto-optimal frontier of intra-layer parallelisation strategies. The search strategy uses linear dynamic programming, but a few steps are required to get there, as it requires the strategy to be formulated as a linear function. Wang et al.~\cite{Wang:2021:ESPL} provide a different search method, focusing on finding a strategy with minimal processing time. Unlike TensorOpt, it does not consider memory usage, optimising just for execution time. 

\section{Conclusion}
\label{sec:conclusion}
Revisiting our research questions, we can conclude the following:

\paragraph{What types of model parallelism exist?}
There are two types of model parallelism: intra-operator, which partitions within an operator, and inter-operator, which partitions over multiple operators. Often, these types are combined into what is referred to as hybrid parallelism, which can also include data-parallelism. 

\paragraph{What are the challenges of model parallelism?}
Challenges include technical trade-offs of the different kinds of model parallelism, with intra-operator having extremely high communication requirements and inter-operator suffering from low device utilisation during training. Finding the optimal parallelisation strategy in hybrid parallelism is another major challenge as the operator-graph and the device-graph most likely will not adequately map onto each other.

\paragraph{What is a modern use-case of model parallelism?}
Model parallelism is currently widely used to train and run inference of multi-billion Transformer models. We find that models from the Megatron family, running on V100 and A100 chips, use intra-operator parallelism within a single compute node and a combination of inter-operator and data parallelism to scale beyond a single node. The PaLM model is able to address the communication challenge of intra-operator parallelism with specialised hardware and does not use any inter-operator parallelism.

\paragraph{Future work}
The field of DNN auto-parallelisation could significantly benefit from standardisation. As discussed in \Cref{sec:discussion}, approaches are often so different that it is impossible to account an advancement in the state-of-the-art to any specific part of the approach. This is in part due to the nature of search problems. However, standardised representations for strategy, device, and model, would help in this regard. For an example of the benefits such standardisation would provide, we can look at other disciplines that deal with search problems,w specifically Neural Architecture Search (NAS). NAS also deals with neural networks and one idea DNN auto-parallelisation could copy from NAS is to provide a data set containing a fully explored search-space, similar to (HW-)NAS-Bench~\cite{Mehta:2022:NASE, Li:2021:HANA}. This would allow methods to be compared without necessitating access to expensive hardware, opening up the field to more people.

%


\balance

\printbibliography

\end{document}